\documentclass[review]{elsarticle}
\usepackage{lineno,hyperref}
\usepackage{amssymb}
\usepackage{amsmath}
\usepackage{multirow}
\usepackage{subfigure}
\usepackage{epstopdf}
\modulolinenumbers[5]

\usepackage{amsmath}
\usepackage{amssymb}
\usepackage{graphicx}

\begin{document}

\begin{frontmatter}
\title{A Dynamic Epidemic Model for Rumor Spread in Multiplex Network with Numerical Analysis
\tnoteref{label1}}
\tnotetext[label1]{This research is supported in part by National Natural Science Foundation of China (No.U181140002 and No.71971031).\\
$^*$The corresponding authors are: dilan@jiangnan.edu.cn (for L.Di);  qguoqi@unimelb.edu.au (for G.Qian) and george$\_$yuan99suda.edu.cn and george$\_$yuan99@yahoo.com (for G.Yuan).}

\author{\textbf{Lan Di$^{a}$,  Yudi Gu$^b$, Guoqi Qian$^{c}$ and George Xianzhi Yuan$^{d,e,f,g}$ }}
\address{$^a$School of Digital Media, Jiangnan University, Wuxi 214122 China}
\address{$^b$Center of Information Construct and Management, Jiangnan University, Wuxi 214122 China}
\address{$^c$School of Mathematics and Statistics, University of Melbourne, Melbourne VIC 3010, Australia}
\address{$^d$Business School, Chengdu University, Chengdu 610106 China}
\address{$^e$School of Financial Technology, Shanghai Lixin University of Accounting and Finance, Shanghai 201209 China}
\address{$^f$Business School and Advanced Institute of Finance, Sun Yat-Sen University, Guangzhou 510275 China}
\address{$^g$Center for Financial Engineering, Sochow University, Suzhou  215031 China}

\begin{abstract}
This paper focuses on studying and understanding of stochastic dynamics in population composition when the
population is subject to rumor spreading.  We undertake the study by first developing an individual
Susceptible-Exposed-Infective-Removed (iSEIR) model, an extension of the SEIR model, for summarizing
rumor-spreading behaviors of interacting groups in the population. With this iSEIR model, the interacting groups
may be regarded as nodes in a multiplex network. Then various properties of the dynamic behaviors of the interacting
groups in rumor spreading can be drawn from samples of the multiplex network. The samples are simulated based
on the iSEIR model with different settings in terms of population scale, population distribution and transfer rate.
Results from the simulation study show that effective control of rumor spreading in the multiplex network entails an
efficient management on information flow, which may be achieved by setting appropriate immunization and spreading
thresholds in individual behavior dynamics. Under the proposed iSEIR model we also have derived a steady-state result,
named the ``supersaturation phenomenon", when the rumor spreading process becomes equilibriumm, which may help us to make the optimal or better control of information flow in the practice.

\end{abstract}


\end{frontmatter}

\nolinenumbers 

\section{Introduction}
\label{sec:introduction}
In this paper we develop a stochastic network model to study rumor spreading dynamics among interacting groups
in a population. Such a model would provide us advanced understanding and insights leading to better strategies
for information management in a communication network. A rumor normally refers to a social communication
phenomenon that can propagate within a human population consisting of interacting groups. Rumors may not
be facts, but they can have significant impact on shaping public opinions, and consequently influencing the
progression of society either positively or negatively. With the help of high-speed internet and abundant use
of various social media, momentum of rumor spreading becomes ever more powerful with regard to both
intensity and rapidity. Therefore, it is important to have an in-depth study of rumor spreading dynamics
in order to properly manage and control rumor spreading for righteous progression of society.
A crucial component of this study is the development of various mathematical models for rumor spreading.

Since the pioneer work of Kermack and McKendrick \cite{Kermack1927}, many mathematical models have been
developed for infectious disease dynamics, leading to the development of many preventive measures and tools to
control or manage infection spread. Among many such models, May and Lloyd \cite{May2001} and Moreno and
Pastor-Satorras et al.\cite{Moreno2002} developed the susceptible-infected-removed (SIR) model and applied it
to analyze the infection spreading in a complex population network.  Infectious disease epidemic and rumor
spreading in complex network actually share many similarities. Therefore, models for infectious disease epidemic
may also be applied to study rumor spreading in principle. In the following we provide a brief review on existent
research for infectious disease epidemic and rumor spreading.

First, based on the classic SIR model, Zhao et al.\cite{Zhao2012} extended the classical SIR model
for rumor spreading by adding a direct link from ignorant to stifler, resulting in a so-called people-Hibernators
model. By relaxing conditions used in previous rumor spreading models, Wang et al.\cite{Wang2014}
developed a new rumor spreading model called SIRaRu, based on which they obtained the threshold of rumor
spreading in both homogeneous and inhomogeneous networks. In addition, through numerical simulations
they found that the underlying network topology exerted significant influence on rumor spreading, and that
the extent of the rumor spreading was greatly impacted by the forgetting rate. Meanwhile, by modelling the
epidemic using a continuous-time Markov chain,  Artalejo et al.\cite{Artalejo2015} developed a
Susceptible-Exposed-Infective-Removed (SEIR) model for quantifying the outbreak duration distribution.
On the other hand, Zhu amd Wang \cite{Zhu2017} proposed a modified SIR model to explore rumor diffusion on
complex social networks, from which they obtained solutions of the corresponding rumor diffusion model.

Second, Granell et al.\cite{Granell2013} introduced a model capable of studying dynamical interplay between
epidemic awareness and spreading in multiplex networks.
Han et al.\cite{Han2014} used the analogy of heat propagation in physics to study the mechanisms and
topological properties of rumor propagation in large-scale social networks, from which they developed a new
model which is shown to have the following peoperties: (1) rumor propagation following this model shall go
through three stages: rapid growth, fluctuant persistence and slow decline; (2) individuals could spread a
rumor repeatedly, so that a resurgence of the rumor is possible; and (3) rumor propagation is greatly influenced
by the rumor's attractiveness, the initial rumormonger and the sending probability.
Considering the possibility of individuals using multiple social networks simultaneously and interactively,
Li et al. \cite{Li2015} proposed a new model for information diffusion in two-layer multiplex networks, by which
they developed a theoretical framework of bond percolation and cascading failure for describing intralayer and
interlayer diffusion. This allowed them to obtain analytical solutions for the fraction of informed individuals as a
function of transmissibility $T$ and interlayer transmission rate $\theta$. Their simulation results showed that
interaction between layers can greatly enhance the information diffusion, and an explosive diffusion is
possible even if the transmissibility of the focal layer is under the critical threshold.
By extending the classical SIRS epidemic model to allow for the infectious forces under intervention strategies
to be governed by a stochastic differential equation (SDE), Cai et al. \cite{Cai2015} used the Markov
semi-group theory to have shown that random fluctuations could suppress disease outbreak, providing us
with a useful control strategy to regulate disease dynamics.

Third, through examining the associations between individuals' behavior and their friends' decisions in a network,
Papagelis et al.\cite{Papagelis2011} used the diffusion dynamics to study the causality between
individual behavior and social influence.
By considering interactions between information awareness and disease spreading, and using the
mean-field theory, Fan et al. \cite{Fan2016} studied the epidemic dynamics and derived the epidemic
thresholds on uncorrelated heterogeneous networks. Their results indicated that interactions between
information awareness and individual behavior influence on the epidemic spreading.
Through numerically examining the interplay between epidemic spreading and awareness diffusion,
Kan and Zhang \cite{Kan2017} showed that the density of the infected and the epidemic
threshold were affected by the two networks and the awareness transmission rate.
This finding was very different from many previous results on single-layer networks: local behavior responses
could alter the epidemic threshold. Moreover, their result indicted that nodes with
more neighbors (hub nodes) in an information network were easier to be informed. Accordingly, the risks of
infection in contact networks could be effectively reduced.

Since all studies aforementioned do not consider the situations where every individual has a
subject-specific probability to become a spreader, we will consider these situations in this paper, for which we develop
a new model to study and understand general rumor spreading behaviors among all interacting groups
in a population. The new model extends the SEIR model and is named
\textit{an individual Susceptible-Exposed-Infective-Removed} (iSEIR) model.
With the iSEIR model we are able to study the distribution of individual behaviors by studying each node
in the corresponding multiplex network. The behaviors distribution can also be numerically simulated
from the iSEIR model with properly specified values of parameters on population scale, population density
and transfer rate, etc.. Our simulation results suggest that the intensity and extensiveness of rumor spreading
can be managed for goodness of society by external intervention.
From the simulation study we also have identified a so-called \textit{supersaturation phenomenon} in rumor
spreading on network, i.e., no individual in the network can be a lurker, which may help us to make the optimal or better control of information flow in the practice.

Contributions of this paper are summarized as following: (i) introducing the iSEIR model capable of describing a
rumor spreading network with individual-specific behaviors over the spread period;
(ii) studying the connecting probabilities, characterized by \textit{population density}, between individuals
belonging to different groups in the network; and
(iii) investigating the dynamic properties of the iSEIR model through a comprehensive simulation study.

In regard to organizing the rest of the paper, review of the related work and the contributions (i) to (iii) are
presented in Sections~\ref{sec2},
\ref{sec3} and \ref{sec4}, respectively. Finally, conclusions and discussions are given in Section~\ref{sec5}.

\section{The Related Work}
\label{sec2}

\subsection{The SIR Model}
\label{sec2.1}

Modeling epidemic spreading starts from a compartmental model, with which the individuals
in the population are divided into groups according to a discrete set of states (e.g., see Murray \cite{Murray1993}).
One such model is the SIR model, cf. Korobeinikov \cite{Korobeinikov2009}, where individuals in the population
are divided into susceptible, infected and removed groups (or states).
Since rumor spreading resembles disease epidemic spreading, it is reasonable to assume an SIR model for
rumor spreading where the three states are replaced by \textit{ignorants}, \textit{spreaders} and \textit{stiflers}.
Denote by $S(t)$, $I(t)$ and $R(t)$ as the proportions of individuals in the populations falling into the three
corresponding states at the time $t$. Also denote by $N$ the population size.
Then for a homogeneous system,  the SIR model can be described by the
following normalization condition
\begin{equation}
  S(t)+I(t)+R(t)=1
\label{eq1}
\end{equation}
and the following system of differential equations:
\begin{equation}
  \left\{
    \begin{aligned}
      \frac{dS}{dt} &=-\mu\langle k\rangle IS \\
      \frac{dI}{dt} &=-\lambda I+\mu\langle k\rangle IS \\
      \frac{dR}{dt} &=\lambda I \\
    \end{aligned}
  \right.
\label{eq2}
\end{equation}
Here $\langle k\rangle$ represents the number of contacts per unit time that is assumed to be constant for
the whole population. In network communication study, $\langle k\rangle$ is interpreted as the average degree
of the network, cf. Wang et al. \cite{Wang2014}. Moreover, quantities $\lambda$ and $\mu$ represent the
removal rate and microscopic spreading (or infection) rate. Equations (\ref{eq1}) and (\ref{eq2}) provide
the following interpretations: (a) Infected individuals decay into the removed class at a rate $\lambda$,
while susceptible individuals become infected at a rate proportional to both the densities of infected and
susceptible individuals, respectively; (b) Under the homogeneous mixing hypothesis used by Murray
\cite{Murray1993}, the force of infection (the per capital rate of acquisition of the disease by the susceptible
individuals) is proportional to the density of infectious individuals.
The homogeneous mixing hypothesis here implies the mean-field treatment to the model, meaning that the rate of
contacts between infectious and susceptible is constant, and independent of any possible source of
heterogeneity present in the system. A further implication from (\ref{eq2}) is that the time scale of the disease is
much smaller than the lifespan of individuals; therefore, we do not need to include in the equation any terms
accounting for the birth or natural death of individuals.

\subsection{The SEIR Model}
\label{sec2.2}

SIR model cannot be applied if susceptible individuals are not immediately infectious after they got infected,
which is the case if the disease involves an incubation period before becoming infectious.
This is resolved by inserting a new state $E$ in between the states $S$ and $I$, resulting in an SEIR model.
For the SEIR model, as is seen in e.g. Bartlett \cite{Bartlett1956}, Allen and Allen \cite{Allen2003} and
De la Sen and Alonso-Quesada \cite{DelaSen2010}, state $S$ refers to
the susceptible group or \textit{ignorants} who are susceptible to disease but have not been infected yet;
state $E$ refers to the exposed group who are infected but are not infectious yet; state $I$ refers to those infected
who also become infectious; and  state $R$ refers to those who have recovered from the infection
(through treatment or natural recovery) and are no longer infectious. We also use $S(t)$, $E(t)$, $I(t)$ and $R(t)$
to represent the proportion of the population being in state $S$, $E$, $I$ and $R$ at time $t$, respectively.

The SEIR model can also be used to describe rumor spreading which shares similar behaviors with the
disease epidemic. In this situation, $(S, E, I, R)$ or $(S(t), E(t), I(t), R(t)$ have the following interpretations:
\begin{enumerate}
\item
$S(t)$ is the proportion of the susceptible (i.e. the ignorant) in the population who do not know the rumor
at time $t$;
\item
$E(t)$ is the proportion of hesitant individuals (i.e. lurkers) who, at time $t$, know the rumor, intend to
but are not yet to spread the rumor;
\item
$I(t)$ is the proportion of those individuals, called spreaders who, at time $t$,  know the rumor
and are also spreading it; and
\item
$R(t)$ is the proportion of those individuals (i.e. stiflers) who know the rumor at time $t$ but are no longer
interest in spreading it.
\end{enumerate}

Based on the work of  De la Sen and Alonso-Quesada \cite{DelaSen2010}, Keeling and Rohani \cite{Keeling2008},
Korobeiniko \cite{Korobeinikov2009},  Kuznetsov and Piccardi \cite{Kuznetsov1994}, Li et al. \cite{Li2001},
Schwartz \cite{Schwartz1983} and the references therein, the SEIR model follows the following ODEs system:
\begin{equation}
  \left\{
    \begin{aligned}
      \frac{dS}{dt} &=-\mu\langle k\rangle SE \\
      \frac{dE}{dt} &=\mu\langle k\rangle SE - \beta EI \\
      \frac{dI}{dt} &=\beta EI - \lambda I \\
      \frac{dR}{dt} &=\lambda I \\
    \end{aligned}
  \right.
\label{eq3}
\end{equation}
where $\mu$ is the infection rate, $\langle k\rangle$ represents the number of contacts per unit time that is
supposed to be constant for the whole population. In network communication language, $\langle k\rangle$ is
interpreted as the average degree of the network, cf. see Wang et al. \cite{Wang2014}. Moreover, $\beta$ is
the rate at which an exposed individual becomes infectious; and $\lambda$ is the recovery rate.
Assume the population is closed with size $N$. Note that, although $I(t)$ has effects on $dS/dt$ in equation system  (3), there is no need to include $I(t)$ explicitly there (and accordingly no need to include $I(t)$ into any equations for $dS/dt$ in the SEIR system). This is because an individual in the $S$ group at time $t$ can only transit to the $E$ group first before possibly transits to the $I$ group after time $t$, cf., the transition diagram given in Figures 1 and 2 below for illustration. Therefore, the effect of $I(t)$ on $dS/dt$ has already been accounted for through including the effect of $E(t)$ on $dS/dt$ at time $t$. By the definitions of $S(t), E(t), I(t)$ and $R(t)$, the SEIR model also meets the normalization condition
\begin{equation}\label{eq4}
  S(t)+E(t)+I(t)+R(t)=1, \quad t\geq0.
\end{equation}

In comparison with the SIR model, the SEIR model gives a more accurate characterization of epidemic
spreading of disease or rumor, if there is an incubation period involved in an individual progressing from
being infected to being infectious. However, the SEIR model does not take into account the variability
in individuals' incubation period, thus may over-estimate the time for the population to become
supersaturated. This gives us motivation to develop an extended SEIR model with individual-specific
behavior in the next section. We also assume that $E(t)$ is not zero throughout the paper in general.

\section{Model for Rumor Spreading with Individual-specific Behaviors}
\label{sec3}

Individual-specific behaviors in disease epidemic have been observed in Rizzo et al.\cite{Rizzo2014} which
lists two such behaviors: one is related to the infected individuals' attempts to suppress the disease spread by
reducing the level of contact with the rest of population; and the other comes from the self-protection of the
susceptible individuals. On the other hand, through studying various activity thresholds in disease epidemic
Liu et al.\cite{Liu2015} found significant effects of the individual-specific behaviors and the transmission network's
topological structure on the spreading dynamics. Significant individual-specific behaviors in rumor spreading
also seem plausible. Thus we will incorporate a probability framework to the SEIR model for modeling such
behaviors in rumor spreading.

\subsection{Framework of individual-specific SEIR model}
\label{sec3.1}

\begin{figure}[htbp]
\centering
  {\includegraphics[scale=0.3]{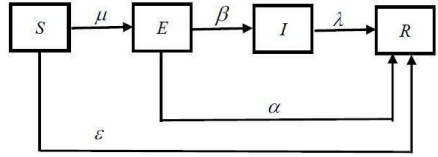}
 \caption{SEIR model with direct transitions to immunity} \label{fig1}}
\end{figure}

Starting with the basic SEIR model  for rumor spreading in a population structured as a multiplex network, we
establish the new model in five steps:

\textbf{Step 1}: We first allow the transition from state $S$ to state $R$ directly with probability $\varepsilon$ per unit time (the same below).
This transition is called \textit{direct immunity} in Chen et al \cite{Chen2011}.
In addition to meeting (\ref{eq4}), the new model satisfies the following ODE system
\begin{equation}\label{f1}
  \left\{
    \begin{aligned}
      \frac{dS}{dt} &=A -\mu SE -\varepsilon S\\
      \frac{dE}{dt} &=\mu SE - \beta EI - \alpha E\\
      \frac{dI}{dt} &=\beta EI - \lambda I \\
      \frac{dR}{dt} &=\varepsilon S + \alpha E + \lambda I \\
    \end{aligned}
  \right.
\end{equation}
where $A$ is the growth rate of new Internet users; $\epsilon$ is the probability of a susceptible person being
directly transformed into an immune person by means of e.g., isolation; $\mu$ is the rate of a susceptible
being infected;
$\beta$ is the rate of an infected person becoming infectious; $\alpha$ is the rate of an infected person becoming
immune directly; and $\lambda$ is the rate of an infectious person entering into an immune state.
Figure~\ref{fig1} gives a visual presentation of (\ref{f1}). Note that there is no direction transition between
$S$ and $I$ in (\ref{f1}).

\textbf{Step 2}: Each individual in the rumor spreading network at time $t$ is identified by its state and position
in that state group. More detail will be given in section~\ref{sec3.2}

\textbf{Step 3}: We will establish an adjacency matrix  to describe the influence effects between individuals
in section~\ref{sec3.2}.

\textbf{Step 4}: Computing the probabilities of transitions between states involves considering the following
two aspects (the $K$-adjacency method):

\textbf{Step 4.1}: the distances between uninfected individuals and their neighborhoods of infected individuals
within; and

\textbf{Step 4.2}: the number of individuals infected.

\begin{figure}[htbp]
\centering
  {\includegraphics[scale=0.3]{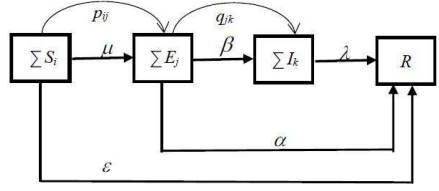}
 \caption{iSEIR model }\label{fig2}}
\end{figure}

\textbf{Step 5}: The full specification of the model is given by combining steps 1 to 4 together with an
individual-level representation of (\ref{f1}) that is illustrated in Figure~\ref{fig2} and to be detailed
in section~\ref{sec3.2}.

The model developed in steps 1 to 5 is named \textit{individual-specific
Susceptible-Exposed-Infective-Removed} (iSEIR) model.

Based on the definitions of $S$, $E$, $I$ and $R$ for the framework of the model described by the equation system (5) (also see (7)), we like to share with readers that for the framework of the model described by equation system (5),  $E$ is a latent population who knows public opinion but has not yet spread, and $I$ is an infectious population who knows public opinion and immediately spreads it. The illustration by Figure 1 (and also Figure 2) explains that a person needs first to change from $S$ to $E$ before becoming $I$, it can't change directly from $S$ to $I$ and thus we do not trade $E$ and $I$ equally in this paper.

\subsection{Individual-level Dynamics Involved in iSEIR}
\label{sec3.2}

The parameters $\varepsilon, \mu, \beta, \alpha$ and $\lambda$ introduced in (\ref{f1}) give the various
population-level effects manifested in the rumor spreading network. These effects can be regarded as
aggregations of the corresponding individual-level effects and cross-individuals effects. We explore the
details in the following.

Recall that $S(t)$ is the proportion of the susceptible in the population at time $t$.
Define $S_i(t)$, $i=1,\cdots, N$, as the probability of individual $i$ being in state $S$ at time $t$.
Then $S(t)=N^{-1}\sum_{i=1}^N S_i(t)$. Similarly we can define $E_i(t), I_i(t)$ and $R_i(t)$ for
$i=1,\cdots, N$. Then $E(t)=N^{-1}\sum_{j=1}^N E_i(t)$, $I(t)=N^{-1}\sum_{k=1}^N I_k(t)$,
and $R(t)=N^{-1}\sum_{l=1}^N R_l(t)$.

In regard to $S$-to-$R$ transition probability $\varepsilon$ per unit time, let us say it is the total of individual-level
$S$-to-$R$ transition contributions. Namely, $\varepsilon =\sum_{i=1}^N \varepsilon_i$.
Similarly let us define $\mu_i$, $\beta_i$, $\alpha_i$ and $\lambda_i$ be the relevant individual-level
transition contributions, $i=1, \cdots, N$. By these definitions and those of $\varepsilon, \mu, \beta, \alpha$
and $\lambda$, we have the following steady-state aggregation equations:
\begin{equation}\label{eq6}
    \begin{aligned}
      \mu &=\sum_{i=1}^N \mu_i; \quad  \varepsilon = \sum_{i=1}^N \varepsilon_i; \quad
\alpha = \sum_{j=1}^N \alpha_j; \quad  \beta = \sum_{j=1}^N \beta_j;  \quad
\lambda = \sum_{k=1}^N \lambda_k
    \end{aligned}
\end{equation}

Since individual-specific transition effects are assumed in our iSEIR model, it is possible that an individual
in one state has influence effect on another individual being in its downstream state. We then define
$p_{ij}$ as the influence effect of individual $i$ in state $S$ on individual $j$ being in state $E$; and
$q_{jk}$ as the influence effect of individual $j$ in state $E$ on individual $k$ being in state $I$.
The respective aggregations of these influence effects are denoted as
$$p=\sum_{i=1}^N\sum_{j=1}^N p_{ij} \quad\mbox{and}\quad
q=\sum_{j=1}^N\sum_{k=1}^N q_{jk}.$$

With all the individual-level quantities aforementioned, the population-level ODE system (\ref{f1}) can be
elaborated into the following individual-level dynamics.

\begin{equation}\label{iSEIR0}
  \left\{
    \begin{aligned}
      \frac{dS}{dt} &=A-S(t)E(t)\sum_{i}\mu_i\sum_{j}p_{ij} - S(t)\sum_{i}\varepsilon_i\\
      \frac{dE}{dt} &=S(t)E(t)\sum_{i}\mu_i\sum_{j}p_{ij} - E(t)I(t)\sum_{j}\beta_j\sum_{k}q_{jk}
- E(t)\sum_{j}\alpha_j\\
      \frac{dI}{dt} &=E(t)I(t)\sum_{j}\beta_j\sum_{k}q_{jk} - I(t)\sum_{k}\lambda_k \\
      \frac{dR}{dt} &=S(t)\sum_{i}\varepsilon_i + E(t)\sum_{j}\alpha_j + I(t)\sum_{k}\lambda_k \\
    \end{aligned}
  \right.
\end{equation}
An illustration of (\ref{iSEIR0}) is given in Figure~\ref{fig2}.

\subsection{Main results}
\label{sec3.3}
In order to present our main results we first need to introduce a concept of \textit{distribution density} $\rho$
which measures the vicinity closeness of individuals in a heterogeneous population.
This concept will also be used in section~4.2 for studying its effect on propagation of rumor spreading.

\noindent
\textbf{Definition 3.1}:  Suppose the population $\mathcal{S}$ for rumor spreading consists of $N$ individuals
$S_j, \; j=1,\cdots, N$; namely $\mathcal{S}=\{S_j, j=1,\cdots, N\}$. Also suppose these $N$ individuals are
distributed over $M$ continuous domains $U_i$, $i=1,\cdots, M$, where a domain may refer to a residential
district or an internet media discussion board. Let $U=\cup_{i=1}^M U_i$, and $C_{i}\in U_{i}$ be the center
of $U_i$ as well as $C\in U$ being the center of $U$. Also let
$\delta(C_{i}, r_0): = \{ y \in U_i: dist(y, C_i) < r_0\}$ be a domain comprising those points
in $U_i$ with their distances to $C_i$ being smaller than $r_0$, and
$\delta(C, r_0): = \{ y \in U: dist(y, C) < r_0\}$ be similarly defined.

Now suppose there exist some minimum radius values $r_1, \cdots, r_M$ and $r$, such that
$U_i\subseteq \delta(C_{i}, r_i), \; i=1,\cdots, M$ and $U\subseteq \delta(C, r)$.
Then the overall vicinity closeness for all individuals in the population $\mathcal{S}$ may be defined
as the distribution density $\rho$:
\[
\rho=\frac{\sum_{i=1}^{M}|\delta(C_{i},r_i)|}{|\delta(C, r)|}
\]
where $|\delta(C_{i},r)|$ is the area  of the domain $\delta(C_{i}, r)$, and $|\delta(C, r)|$ is similarly defined.
Note that 1): $\rho=0$ implies that $M=N$, each $S_i$ is the center, and $r_i=0$.
Thus $S_1, \cdots, S_N$ are uniformly distributed over $\mathcal{S}$; and 2): $\rho=1$ implies $M=1$
and $r_1=r$.

Now according to Gonalez-Parra et al.\cite{Gonalez2010} and by the fact that the first three equations in the
model (\ref{iSEIR0}) do not contain the variable $R(t)$, we can conclude that the dynamics in (\ref{iSEIR0})
can be completely represented at the population-level by the first three equations 	
\begin{equation}\label{iSEI2}
  \left\{
    \begin{aligned}
      \frac{dS}{dt} &=A-\mu pS(t)E(t)- \varepsilon S(t) \\
      \frac{dE}{dt} &=\mu pS(t)E(t) - \beta qE(t)I(t) - \alpha E(t) \\
      \frac{dI}{dt} &=\beta qE(t)I(t) - \lambda I(t) \\
    \end{aligned}
  \right.
\end{equation}

Now based on the propagation dynamics theory introduced in e.g. Zhao et al \cite{Zhao2012}), we know that
the behavor of the whole rumor spreading system depends on certain propagtion threshold parameter $R_0$
(which is also called the basic regeneration number). In particular, $R_0$ has impact on the equilibrium
distribution of rumor spreading states. Specifically,  (1): when  $R_0\leq1$, the rumor spread will eventually
disappear; and (2): when $R_0 > 1$, the rumor spreading will achieve to an equilibrium distribution.
These properties will be confirmed by Theorem~3.3 later in this section.

But we first follow van den Driessche and Watmough \cite{Van2002} to obtain an expression for $R_0$.
Denoting $\mathbf{x}: =(E, I, S)^T$, the model system (\ref{iSEI2}) can be expressed as
 $$\frac{d\mathbf{x}}{dt}: = F(\mathbf{x}) - V(\mathbf{x}),$$
where
\begin{equation}\label{fx}
  F(\mathbf{x})=\left (
    \begin{aligned}
      \mu pSE \\
      0\\
      0\\
      \end{aligned}
  \right ).
\end{equation}
\begin{equation}\label{vx}
  V(\mathbf{x})=\left (
    \begin{aligned}
     \beta qEI+\alpha E \\
     -\beta qEI+\lambda I \\
      -A+\mu pSE+\varepsilon S \\
      \end{aligned}
  \right ).
\end{equation}

By defining $G: = FV^T$, the available spectral radius (i.e., the basic regeneration number $R_0$) can be
found from van den Driessche and Watmough \cite{Van2002} to be
\begin{equation}\label{R0}
R_0: = \xi(G) = \frac{(\beta q)(\mu p)}{(\beta q+\alpha)\lambda}
\end{equation}

A plausible initial setting is needed for studying the dynamics of rumor spreading. For this we assume there is
only one spreader at the beginning, and the initial setting for rumor spreading is given by
$$S(0) =\frac{N-1}{N} , E(0)=\frac{1}{2N} , I(0)=\frac{1}{2N} , R(0)=0.$$

Next we provide two lemmas which are taken from Zhao et al. \cite{Zhao2012}

\textbf{Lemma 3.1}: For $\nu >1$, equation $R =1- e^{-\nu R}$ has two solutions of $R$: a trivial one $R=0$
and a nontrivial one $0<R<1$.

\textbf{Proof}: It is Theorem 1 of Zhao et al. \cite{Zhao2012}, which completes the proof. \hfill $\square$

\textbf{Lemma 3.2}: For equation $R= 1- e^{-\epsilon R}$, where $\epsilon = \frac{\lambda + \alpha}{\alpha}$,
we have that for a fixed $\alpha$, $R$ increases as $\lambda$ increases. Similarly, given a fixed $\lambda$,
$R$ decreases as $\alpha$ increases.

\textbf{Proof}: It is Theorem 2 of Zhao et al. \cite{Zhao2012}, which completes the proof. \hfill $\square$

In the following we aim to establish a general theoretic result for final removal proportion, to be presented in
Theorem~3.3, for rumor spreading that follows the iSEIR model. Here the final removal proportion in rumor
spreading dynamics is defined as
$$ R: = \mbox{Final}\{R(t)\} = \lim_{t\to\infty} R(t) = R(\infty), $$
which can be used to measure the level of rumor influence in practice.

When the dynamics of rumor spreading following the iSEIR model eventually achieves equilibrium,
it is reasonable to assume that $A\approx0$, $\varepsilon$ is close to $0$, $I\approx E$ (i.e. the proportion
of the infected being lurkers is nearly zero), and network size $N$ is sufficiently large.
With these assumptions, we have the following key result.

\textbf{Theorem 3.3}: Let $\nu=\frac{\mu p}{\alpha + \lambda}$. Then when
$\mu p > \alpha + \lambda$, the equation  $R=1- e^{-\nu R}$ has two solutions:
zero solution and a nontrivial solution $R$ satisfying $ 0 < R < 1$.

\textbf{Proof}: Based on the system of equations (\ref{f1}) and (\ref{iSEI2}), we have
\begin{equation}\label{11}
\frac{dR}{dS}= \frac{\varepsilon S+\alpha E+\lambda I}{A-\mu pSE-\varepsilon S}.
\end{equation}
Assuming $A\approx0$,  we have
\begin{equation}\label{12}
dR = \frac{\varepsilon S+\alpha E+\lambda I}{-\mu pSE-\varepsilon S}dS
\end{equation}
Now integrating both sides of Eq.(\ref{12}) from the initial time to the stationary time and noting that $\varepsilon$
is close to $0$, it follows that
\begin{equation}
\int^{\infty}_{0}dR=\int^{\infty}_{0}\frac{\alpha E+\lambda I}{-\mu pSE}dS.
\end{equation}
Then
\begin{equation}
R(\infty)-R(0)=\frac{\alpha E+\lambda I}{-\mu pE}[\ln(S(\infty))-\ln(S(0))].
\end{equation}

Noting that $S(0)=\frac{N-1}{N}\approx1 ( \mbox{as}\; N \to \infty)$, $R(0)=0$, $S(\infty) = 1- R(\infty)= 1-R $,
and $R(\infty)=R$, thus we have
\begin{equation}\label{13}
R = -\frac{\alpha E+\lambda I}{\mu pE}\ln(1-R)
\end{equation}

\begin{equation}\label{14}
-\frac{\mu pR}{\alpha+\lambda (I/E)}=\ln(1-R).
\end{equation}

From Eq.(\ref{14}) it follows that
\begin{equation}\label{}
e^{-\frac{\mu pR}{\alpha+\lambda (I/E)}}=1-R.
\end{equation}
Thus we obtain the following transcendental equation
\begin{equation}\label{}
R = 1-e^{-\frac{\mu p}{\alpha+\lambda(I/E)}R}.
\end{equation}
By the assumption that $I\approx E$, it follows that
\begin{equation}\label{}
R = 1-e^{-\frac{\mu p}{\alpha+\lambda}R}.
\end{equation}
Now by applying Lemma 3.1 above, let $\nu=\frac{\mu p}{\alpha + \lambda}$, we have $\nu> 1$, this
implies that the conclusion is true, which completes the proof. \hfill $\square$

Theorem~3.3 gives an equation that must be satisfied by the steady-state removal proportion $R$ in the
rumor spreading dynamics that follow the iSEIR model. This provides guidance to conducting numerical
simulations to be given in Section~4, where the supersaturation phenomenon can be observed in
rumor spreading dynamics if ``lurkers" do not exist in the network.

\section{Numerical Simulation and Analysis}
\label{sec4}
In this section we will present three simulation studies based on the developed iSEIR model, then summarize
the results. The goal is to improve our understanding and develop insights on the effects of the population
size, the individual distribution density, and transition probabilities among various states on the rumor propagation
dynamics.

In our simulation experiments, unless specified otherwise the number of domains $M$ for the population
is set to be $10$, the size of population is $N=400$, and each experiment for the given $M$ and $N$ is
repeated $100$ times to complete a simulation.
In each simulation, we use the Euler algorithm to generate the proportions $\{(S(t), E(t), I(t), R(t)):
t=0,1,\cdots, T\}$ from the iSEIR model (\ref{iSEIR0}), with $T=2,000$ being set unless specified
otherwise; and the time unit used is 5-minute so $T=2,000$ corresponds to about 7 days.
Note that the simulation underlying Figure 4 in Session 4.1 chooses $T=5,000$ (corresponding to 17.5 days)
and $N=800\;\mbox{and}\; 10,000$.

Values of all parameters used in (\ref{iSEIR0}) in the simulations are generated according to instructions
listed in Table~\ref{Table1}, where $d(i,j)$ is the distance between points $i$ and $j$, and \textit{rnd}
is a random number following Uniform(0,1) distribution. The quantity $c$ in Table~\ref{Table1} is a
distance threshold:
\begin{equation}\label{}
c = \frac{1}{N^2}\sum_{i}^N\sum_{j}^Nd(i,j)
\end{equation}

\subsection{Influence of Population size $N$ on Rumor Propagation Dynamics}
\label{sec4.1}

\begin{figure}[htbp]
\setlength{\abovecaptionskip}{0.cm}
\setlength{\belowcaptionskip}{-0.cm}
\vfill{}
\subfigure[$N$=100]
{\includegraphics[scale=1.0]{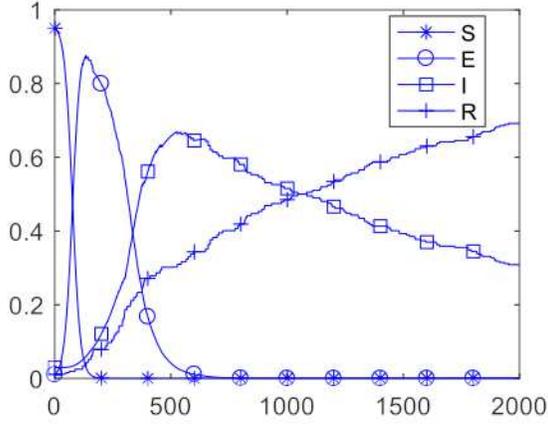}
}
\subfigure[$N$=200]
 {\includegraphics[scale=1.0]{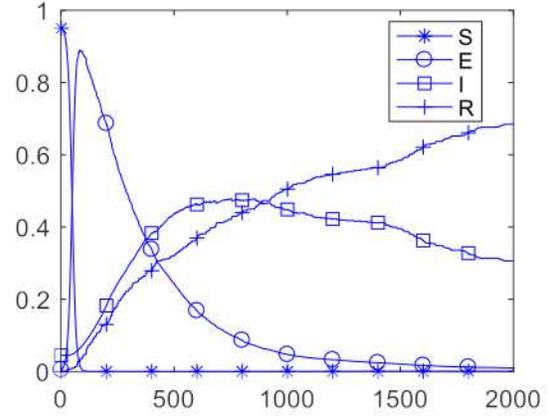}
 }
\vfill{}
\subfigure[$N$=400]
 {\includegraphics[scale=1.0]{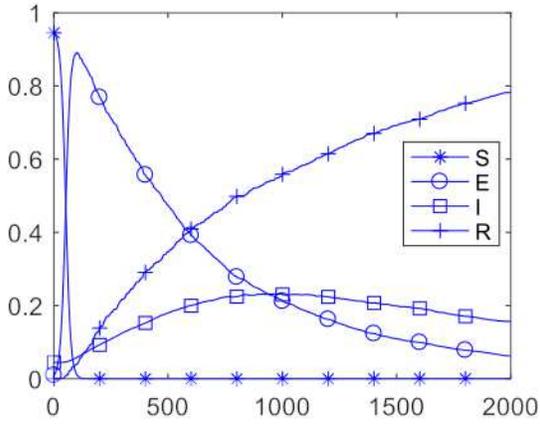}
}
\subfigure[$N$=800]
 {\includegraphics[scale=1.0]{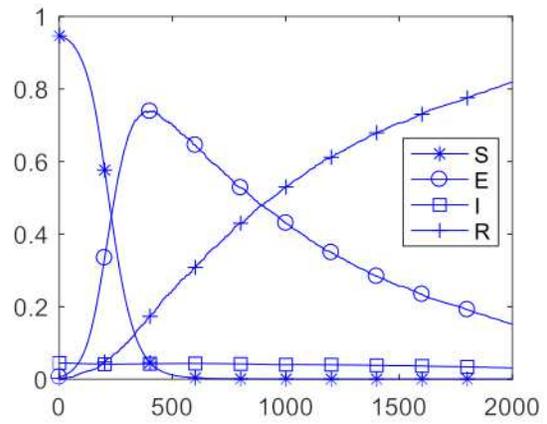}
}
\protect\caption{Rumor spreading performance from a typical experiment}\label{fig3}

\end{figure}

In this subsection we assess the effect of populaton size $N$ on rumor spreading dynamics.
We set $N: = 100$, $200$, $400$ and $800$,  respectively. We also set $M=10$ and repeat the experiment
100 times. Noting that the rumor spreading dynamics vary from experiment to experiment due to
individual-specific behavors involved, we display in Figure~\ref{fig3} the performance of only a typical
experiment.

\setlength{\tabcolsep}{4pt}
\begin{table}
\small
\begin{center}
\caption{
Parameters set used in experiments\label{Table1}}
\begin{tabular} {ccc}
\hline\noalign{\smallskip}
value &  1  &	0 \\
\hline
The Parameter & \multicolumn{2}{c}{The Criteria Condition}\\
\hline
$p_{ij}$	& $d(i,j)<c$ &otherwise\\
$q_{jk}$	& $d(j,k)<c$ &otherwise\\
$\mu_i$	    & $rnd\in [0.0001,1]$	&otherwise\\
$\varepsilon_i$	&$rnd<0.0001$	&othewise\\
$\beta_j$	& $rnd\in [0.001,1]$	&otherwise\\
$\alpha_j$	&$rnd<0.001$	&otherwise\\
$\lambda_j$ &$rnd<0.0005$	&otherwise\\
\hline
\end{tabular}
\end{center}
\end{table}

From (a) to (d) in Figure~\ref{fig3} we can observe that:
1) as $N$ increases, all curves are getting smoother and smoother; and 2) as $N$ increases, less and less
individuals stay in state $I$ at any given time $t$. For example $I(t)<0.1$ when $N=800$ and
$t$ is sufficiently large.

On the other hand, it has been observed that variation in rumor spreading dynamics among all 100 experiments
becomes smaller and smaller when population size $N$ increases. Indeed our simulation results show that:
1) when $N=100$ and $200$, more than $30\%$ of the repeated experiments show similar behaviors
as shown in Fig.\ref{fig3}(a) and Fig.\ref{fig3}(b); 2) when $N=400$, more than $40\%$ of the repeated
experiments show similar behaviors as shown in Fig.\ref{fig3}(c); and 4) when $N=800$, we observed that more than $90\%$ of the repeated experiments show similar behaviors as given in Fig.\ref{fig3}(d).

\begin{figure}[htbp]
\setlength{\abovecaptionskip}{0.cm}
\setlength{\belowcaptionskip}{-0.cm}
\vfill{}
\subfigure[$N$=800]
{\includegraphics[scale=1.0]{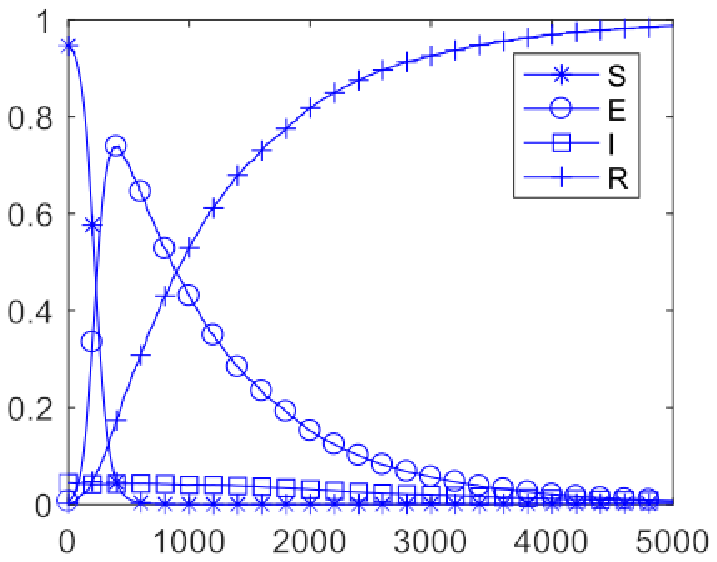}
}
\subfigure[$N$=10000]
 {\includegraphics[scale=1.0]{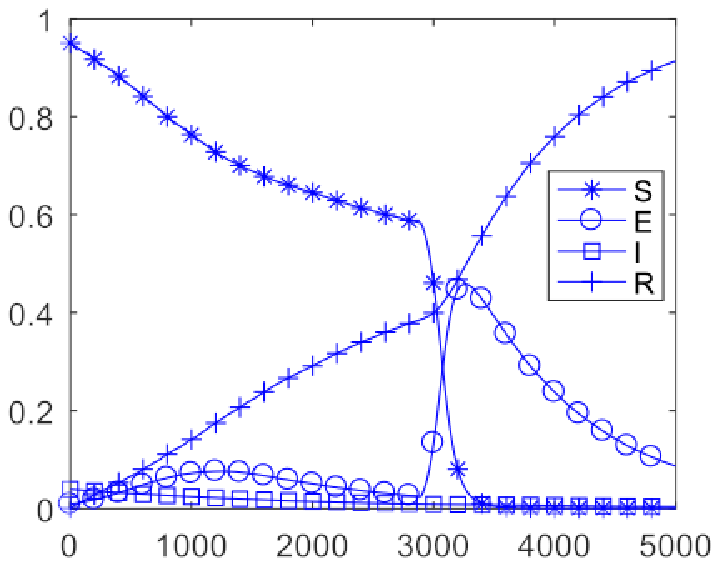}
}
\protect\caption{Case of large-scale population data}\label{fig4}
\end{figure}

We also have simulated the rumor spreading dynamics in 100 repeated experiments with an increased $T$ of
5,000 (equivalent to 17.5 days) and $N$ of 800 and 10,000. The typical performance is displayed in
Figure~\ref{fig4}.

From Figure~\ref{fig4}, we observe that rumor spreading in terms of $(S(t), E(t), I(t)$, $R(t))$ dynamics becomes
stationary after about 4,500 time units when the population size $N$ is not large. However, when $N$ is large,
the rumor spreading has a markedly different pattern. In particular, the exposure proportion $E(t)$ along with
the time goes upward first, then goes downward, then goes upward again and sharply before goes downward.
Around $t=3000$ (equivalent to 10.5 days) seems to be a critical moment when $E(t)$ changes from
downward to sharp upward.
The infectiousness proportion $I(t)$ and the susceptible proportion $S(t)$ both behave similarly to when $N$
is not large, with $I(t)<0.05$ and $S(t)<0.05$ when $t$ is sufficiently large. Due to the significant change of
behavior of $E(t)$ observed, the behavior of the removal proportion $R(t)$ is moderately different when $N$
is large from when $N$ is not large. Nevertheless, this change of behavior in $R(t)$ is still within the
expectation specified in Theorem~3.3. Actually, from the parameter setup used for generating the dynamics
presented in Figure~\ref{fig4}(b), we have obtained $c=0.3$, $p=0.0222$, $\mu=0.999$, $\alpha=0.00099$
and $\lambda=0.00049$. This shows that $p\mu = 0.0022$ which is greater than $0.00149 (=\alpha + \lambda)$.
Therefore, result of Theorem~3.3 applies in the current setup.

\subsection{Influence of Individual Distribution Density on Rumor Propagation}
\label{sec4.2}

In section~\ref{sec4.1} we assume all individuals in the population distribute uniformly over its domain $U$,
i.e. we assume the distribution density (or concentration) $\rho=0$ with $\rho$ being defined in
section~\ref{sec3.3}. Now we would like to see by simulation the behavior of rumor spreading in the network when
the individuals do not distribute uniformly, i.e. when $0<\rho\leq 1$.

\begin{figure}[htbp]
\vfill{}
\subfigure[Similar figures appear 43\%.]
{\includegraphics[scale=0.95]{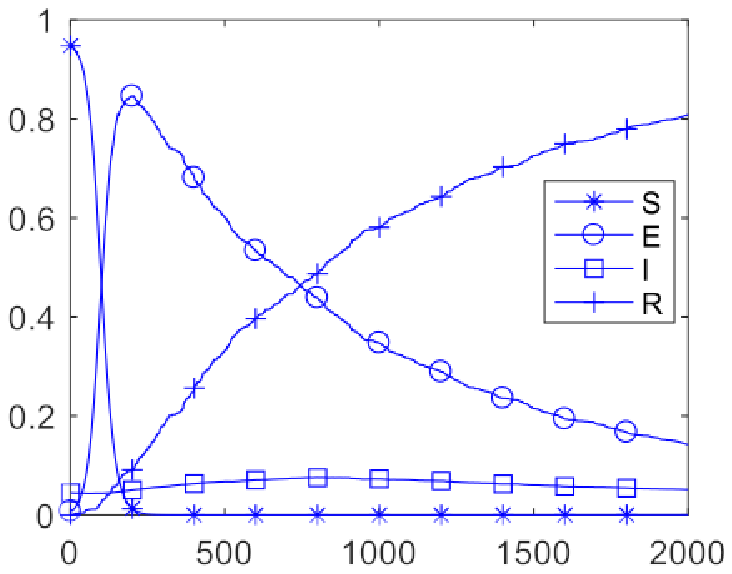}
}
\subfigure[Similar figures appear 25\%.]
 {\includegraphics[scale=0.95]{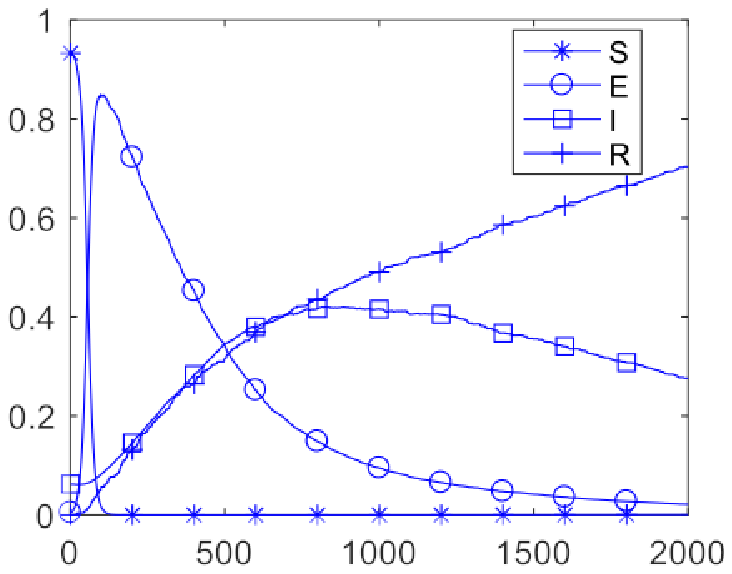}\\
}
\protect\caption{Propagation with population concentration($\rho=0$)}\label{fig5}
\end{figure}

\begin{figure}[htbp]
\setlength{\abovecaptionskip}{0.cm}
\setlength{\belowcaptionskip}{-0.cm}
\vfill{}
\subfigure[Similar figures appear 46\%.]
{\includegraphics[scale=0.95]{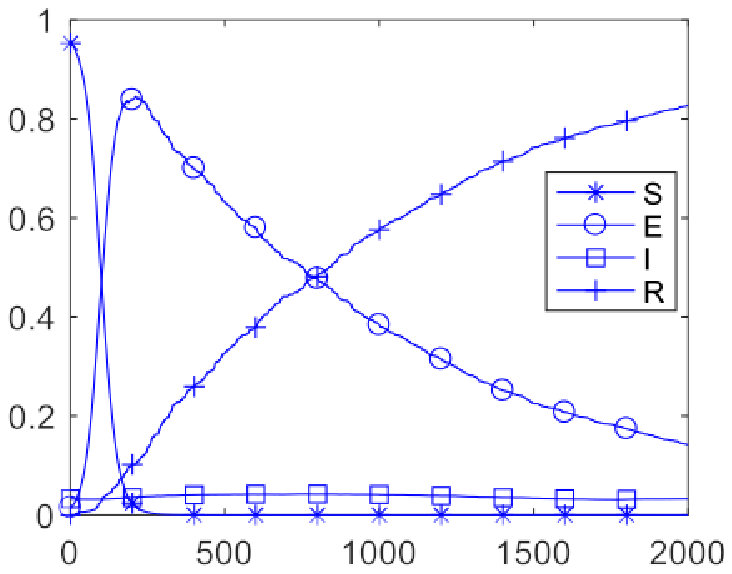}
}
\subfigure[Similar figures appear 31\%.]
 {\includegraphics[scale=0.95]{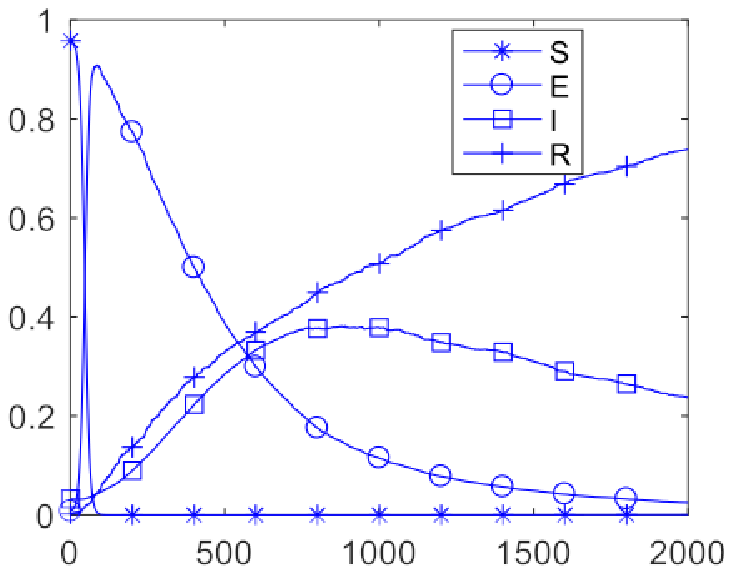}
}
\protect\caption{Propagation with population concentration($\rho=0.2$)}\label{fig6}
\end{figure}

\begin{figure}[htbp]
\setlength{\abovecaptionskip}{0.cm}
\setlength{\belowcaptionskip}{-0.cm}
\vfill{}
\subfigure[Similar figures appear 50\%.]
{\includegraphics[scale=0.95]{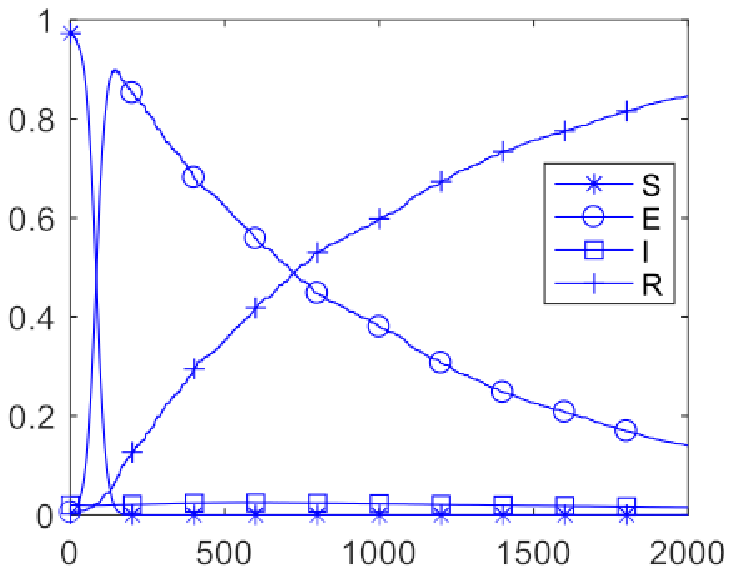}
}
\subfigure[Similar figures appear 28\%.]
 {\includegraphics[scale=0.95]{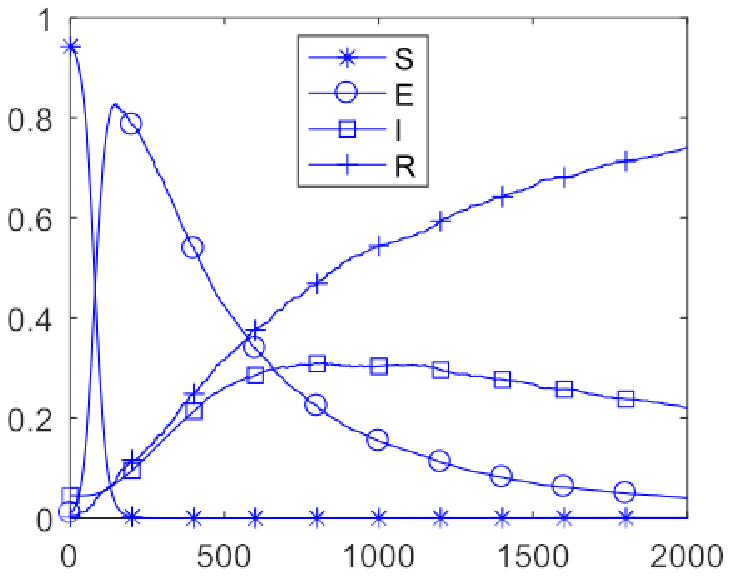}
}
\protect\caption{Propagation with population concentration($\rho=0.4$)}\label{fig7}
\end{figure}

\begin{figure}[htbp]
\setlength{\abovecaptionskip}{0.cm}
\setlength{\belowcaptionskip}{-0.cm}
\vfill{}
\subfigure[Similar figures appear 57\%.]
{\includegraphics[scale=0.95]{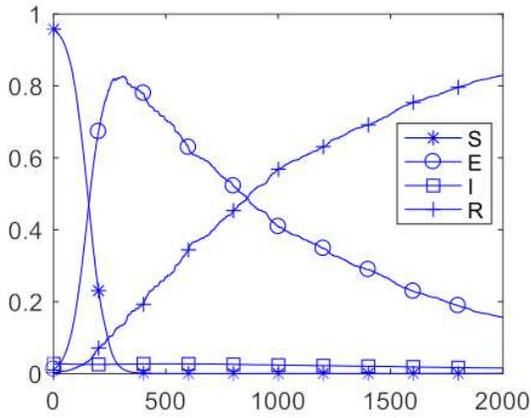}
}
\subfigure[Similar figures appear 23\%.]
 {\includegraphics[scale=0.95]{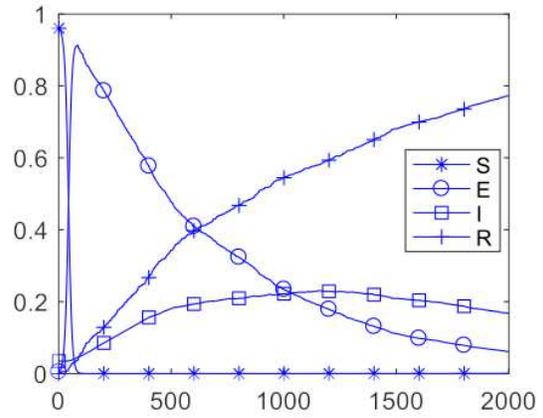}
}
\protect\caption{Propagation with population concentration($\rho=0.6$)}\label{fig8}
\end{figure}

In this simulation we set $M=10$ and $N=400$, and each experiment is repeated 100 times. Due to
non-uniform distribution of individuals in the domain, we need the influence effect probabilities
$p_{ij}$'s and $q_{jk}$'s defined in section~\ref{sec3.2} for generating $S(t)$, $E(t)$, $I(t)$ and $R(t)$
values from the iSEIR model. The $p_{ij}$ and $q_{jk}$ values are to be specified according to
Table~\ref{Table1}, from which we can see they are dependent on whether $d(i,j)<c$ and $d(j,k)<c$ or not.
Accordingly, $p_{ij}$ and $q_{jk}$ values are related to $\rho$.
Typical behaviors of rumor spreading for this setup are displayed in Figures~\ref{fig5} to \ref{fig8}, where
$\rho=0$, 0.2, 0.4 and 0.6 respectively. In each of captions of these eight figures, there is a percentage
number which is for summarizing the overall rumor propagation. For example, refer to Figure~\ref{fig8}(a)
and (b) we can say that when $\rho=0.6$, 57\% of the 100 experiments have performance similar to that
in Figure~\ref{fig8}(a), and 23\% of them have performance similar to that in Figure~\ref{fig8}(b).

From Figure~\ref{fig5}(a) to Figure~\ref{fig8}(b) we have the following observations.
\begin{enumerate}
\item
When the distribution concentration density is $\rho=0$, in 43\% of the experiments the infectiousness
proportion $I(t)$ achieves equilibrium from $t=1500$ on; while in 25\% of the experiments $I(t)$ still has not
achieved equilibrium by time $t=2000$.
\item
When the distribution concentration density $\rho$ departs from 0 and increases to 0.6, the rumor spreading
dynamic system are still stable eventually. The percentage of the experiments where $I(t)$ become well
stabilized (i.e. $I(t)<0.1$) gradually increases from 43\% to 57\%. The percentage for where $I(t)$ is not
yet stable fluctuates but has a trend of decreasing. It is interesting to see that high distribution concentration
density somewhat suppresses the rate of rumor spreading sometimes. It may be interpreted that during
these situations many of the individuals move from the \textit{exposure} state to the \textit{removal} state
directly rather than move to the \textit{infectiousness} state first.
\item
The property established in Theorem~3.3 is still applicable in the simulations underlying Figures~\ref{fig5}(a)
to \ref{fig8}(b), no matter whether $\rho=0$ or $\rho>0$.
\end{enumerate}

From the above observations we see that rumor spreading is likely to decelerate due to increase in
individuals distribution density. It is possible that many infected individuals (i.e. those in state \textit{exposure})
will skip state \textit{infectious} and move to state \textit{removal} directly, resulting in the so-called
\textit{supersaturation} phenomenon. Those individuals who have exposure to the rumor but do not
actually spread the rumor may be referred to the \textit{lurkers}.

\subsection{Effect of Infectiousness to Removal Transition on Rumor Propagation}
\label{sec4.3}

\begin{figure}[htb!]
\setlength{\abovecaptionskip}{0.cm}
\setlength{\belowcaptionskip}{-0.cm}
\vfill{}
\subfigure[Similar figures appear 59\%.]
{\includegraphics[scale=0.95]{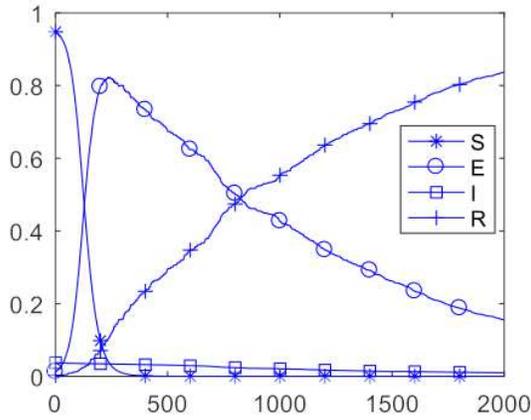}
}
\subfigure[Similar figures appear 27\%.]
 {\includegraphics[scale=0.95]{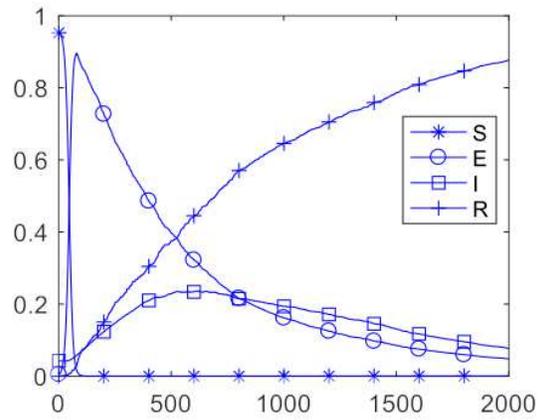}
}
\protect\caption{The Propagation with different transfer rate.$\lambda=0.001$}\label{fig9}
\end{figure}

\begin{figure}[htb!]
\setlength{\abovecaptionskip}{0.cm}
\setlength{\belowcaptionskip}{-0.cm}
\vfill{}
\subfigure[Similar figures appear 50\%.]
{\includegraphics[scale=0.95]{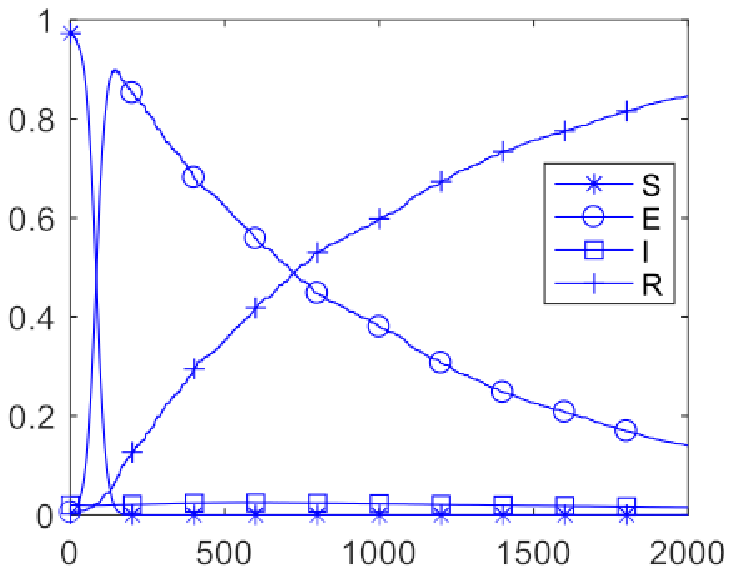}
}
\subfigure[Similar figures appear 28\%.]
 {\includegraphics[scale=0.95]{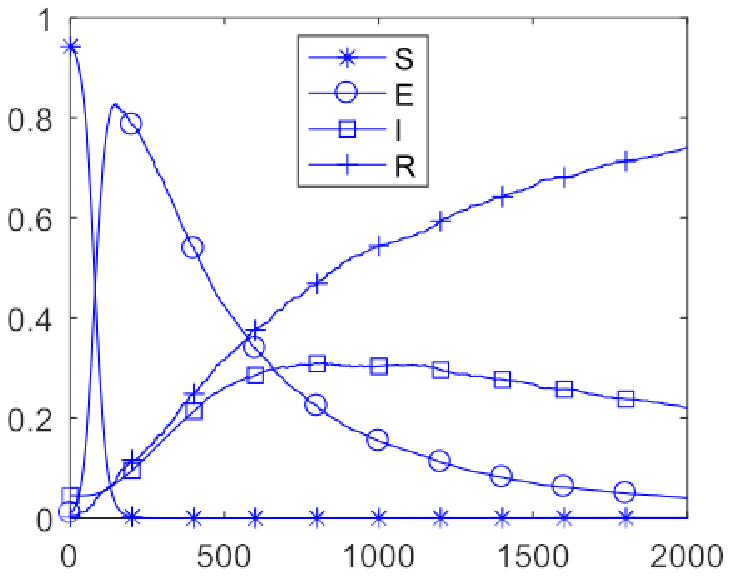}
}
\protect\caption{The Propagation with different transfer rate.$\lambda=0.0005$}\label{fig10}
\end{figure}

\begin{figure}[htb!]
\setlength{\abovecaptionskip}{0.cm}
\setlength{\belowcaptionskip}{-0.cm}
\vfill{}
\subfigure[Similar figures appear 42\%.]
{\includegraphics[scale=0.95]{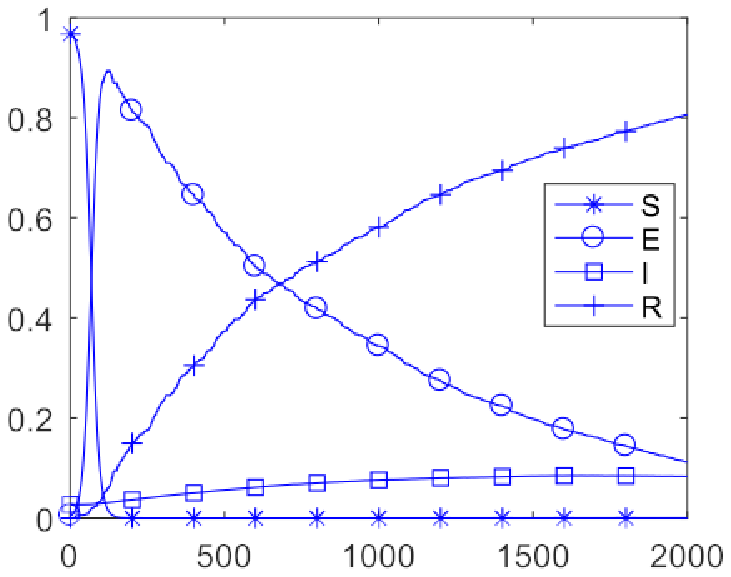}
}
\subfigure[Similar figures appear 30\%.]
 {\includegraphics[scale=0.95]{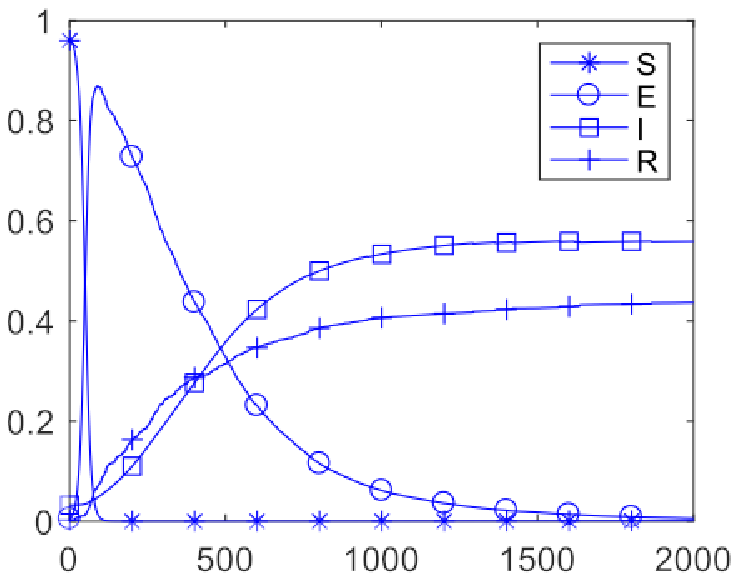}
}
\protect\caption{The Propagation with different transfer rate.$\lambda=0.0001$}\label{fig11}
\end{figure}

\begin{figure}[htb!]
\setlength{\abovecaptionskip}{0.cm}
\setlength{\belowcaptionskip}{-0.cm}
\vfill{}
\subfigure[Similar figures appear 30\%.]
{\includegraphics[scale=0.95]{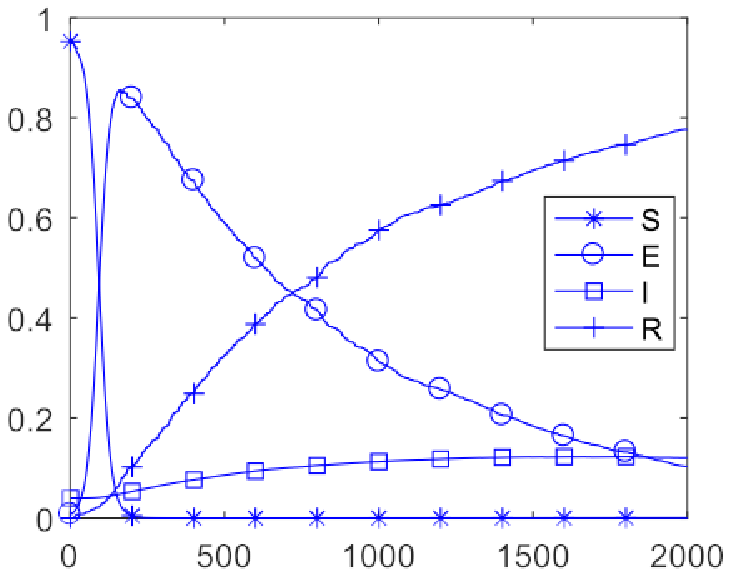}
}
\subfigure[Similar figures appear 33\%.]
 {\includegraphics[scale=0.95]{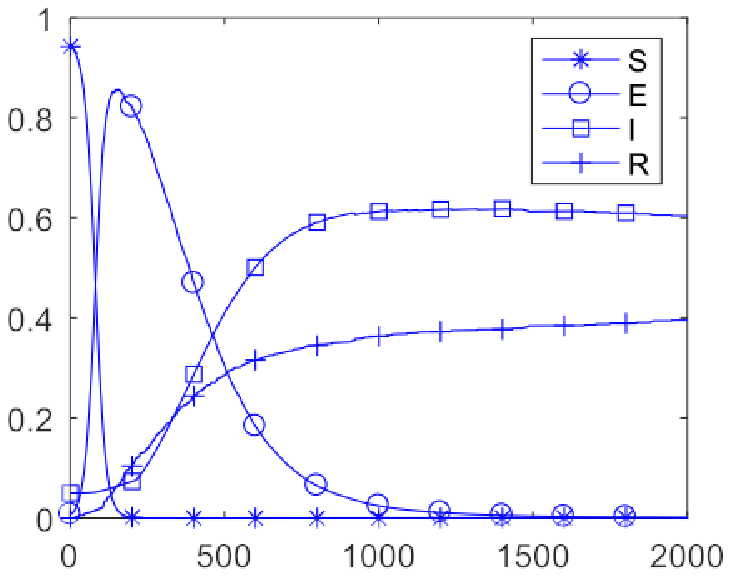}
}
\protect\caption{The Propagation with different transfer rate.$\lambda=0.00005$}\label{fig12}
\end{figure}

In this subsection we use simulation to study how the rumor spreading dynamics will vary if the transition
probability $\lambda$ of individuals in the population moving from the infectiousness state to the removal
(i.e. immune) state varies.
Here we set $\lambda$ to $0.001$, $0.0005$, $0.0001$ and $0.00005$, respectively. Setup of the
other parameters in the simulation remains the same as in previous subsections, i.e. $M=10$, $N=400$
and the experiment is repeated 100 times.
Typical performances in the simulation are displayed in Figures~\ref{fig9}(a) to \ref{fig12}(b).

We have the following observations from these figures.
\begin{enumerate}
\item
Left column of plots show the cases when the infectiousness proportion $I(t)$ eventually gets controlled
under 0.1. In these cases the rate of $I(t)$ going below 0.1 along the timeline decreases as $\lambda$
decreases from 0.001, 0.0005, 0.0001 to 0.00005. The proportion of experiments showing this behavior
decreases from 59\%, 50\%, 42\% to 30\%, however. It implies that, the proportion $I(t)$ gradually
is more and more likely to be out of control (i.e $> 0.1$) as time goes.
\item
Right column of plots show the cases when the infectiousness proportion $I(t)$ still has not been under
control ($<0.1$) by time $t=2000$. In these cases, $I(t)$ is larger and larger, i.e. increases from
0.2 to 0.6 when $\lambda$ decreases.
The proportion of experiments showing this behavior increases from 27\%, 28\%, 30\% to 33\% when
$\lambda$ decreases from 0.001, 0.0005, 0.0001 to 0.00005.
This performance conforms to the definition of $\lambda$ that is the transition probability of an individual
moving from infectiousness to removal. It also conforms to the supersaturation phenomenon.
\end{enumerate}

\section{Discussion and Conclusion}
\label{sec5}
This paper is motivated by the desire to understanding the rumor propagation dynamics in a population of
individuals from both the population and the individual levels. We have developed an iSEIR model for
studying this dynamics. The iSEIR model substantially extends the classical SIR model by introducing
an exposure state and an individual-specific transition framework. While most SIR related research works
focus on public health and epidemiology, we apply the new iSEIR model in the context of rumor spreading
dynamic system which we believe have produced innovative and important results to research in
social network study.

In addition to developing general theoretic results, we have performed three simulation studies to investigate
the effects of populations size, population distribution concentration density and infectiousness-to-removal
transition probability on the behaviors of rumor spreading dynamics. Our simulation studies have
produced some interesting observations, e.g. the supersaturation phenomenon in which the infectiousness
proportion $I(t)$ may not grow quickly at any time but it persists to very long, especially when
the distribution density is high or $\lambda$ is small. Another observation is individuals in a population
with large size tend to have more stable potential to influence their neighbors.

Although we have obtained a number of interesting results for rumor spreading dynamics, we would
like to point out that further works are required to improve understanding of the individual-specific effects
on rumor spreading dynamics. This should be delegated to our future research.

\section*{Acknowledgement}
This research is supported in part by the National Natural Science Foundation of China (No. U181140002).
%



\end{document}